\documentclass{aa}
\usepackage{amssymb}
\usepackage{graphics}
\newcommand {\ic} {\'{\i}}
\begin{document}
%
%
   \thesaurus{03        
             (13.07.1;  
              13.07.2;  
              09.03.2)} 
%
%
   \title{Search for gamma-ray bursts above 20 TeV with the HEGRA AIROBICC
Cherenkov array}
%
%
   \author{L. Padilla         \inst{1}
      \and B. Funk            \inst{2}
      \and H. Krawczynski     \inst{3,4}
      \and J.L. Contreras     \inst{1}
      \and A. Moralejo        \inst{1}
      \and F. Aharonian       \inst{3}
      \and A.G. Akhperjanian  \inst{5}
      \and J.A. Barrio        \inst{1,6}
      \and J.G. Beteta        \inst{1}
      \and J. Cortina         \inst{1}
      \and T. Deckers         \inst{7}
      \and V. Fonseca         \inst{1}
      \and H.-J. Gebauer      \inst{6}
      \and J.C. Gonz\'alez    \inst{1,6}
      \and G. Heinzelmann     \inst{4}
      \and D. Horns           \inst{4}
      \and H. Kornmayer       \inst{6}
      \and A. Lindner         \inst{4}
      \and E. Lorenz          \inst{6}
      \and N. Magnussen       \inst{2}
      \and H. Meyer           \inst{2}
      \and R. Mirzoyan        \inst{5,6}
      \and D. Petry           \inst{2,6}
      \and R. Plaga           \inst{6}
      \and J. Prahl           \inst{4}
      \and C. Prosch          \inst{6}
      \and G. Rauterberg      \inst{7}
      \and W. Rhode           \inst{2}
      \and A. R\"ohring       \inst{4}
      \and V. Sahakian        \inst{5}
      \and M. Samorski        \inst{7}
      \and D. Schmele         \inst{4}
      \and W. Stamm           \inst{7}
      \and B. Wiebel-Sooth    \inst{2}
      \and M. Willmer         \inst{7}
      \and W. Wittek          \inst{6}
   }
   \offprints{padilla@gae.ucm.es}
   \institute{Facultad de Ciencias F\ic sicas, Universidad Complutense, E-28040
              Madrid, Spain
      \and
              BUGH Wuppertal, Fachbereich Physik, Gau\ss str.20, D-42119
              Wuppertal, Germany
      \and
              Max-Planck-Institut f\"ur Kernphysik, P.O. Box 103980, D-69029
              Heidelberg, Germany
      \and
              Universit\"at Hamburg, II. Inst. f\"ur Experimentalphysik,
              Luruper Chausee 149, D-22761 Hamburg, Germany
      \and
              Yerevan Physics Institute, Yerevan, Armenia
      \and
              Max-Planck-Institut f\"ur Physik, F\"ohringer Ring 6, D-80805
              M\"unchen, Germany
      \and
              Universit\"at Kiel, Institut f\"ur Experimentelle und Angewandte
              Physik, D-24098 Kiel, Germany
   }
   \date{Received XX Xxxxxxxxx 1998 / Accepted XX Xxxxxxxxx 1998}
%
   \maketitle
%
%
   \begin{abstract}
   A search for gamma-ray bursts (GRBs) above 20 TeV within the field of view
   (1 sr) of the HEGRA AIROBICC Cherenkov array (29$^\circ$N, 18$^\circ$W, 2200
   m a.s.l.) has been performed using data taken between March 1992 and March
   1993. The search is based on an all-sky survey using four time scales, 10
   seconds, 1 minute, 4 minutes and 1 hour. No evidence for TeV-emission has
   been found for the data sample. Flux upper limits are given. A special
   analysis has been performed for GRBs detected by BATSE and WATCH. Two
   partially and two fully contained GRBs in our field of view (FOV) were
   studied. For GRB 920925c which was fully contained in our FOV, the most
   significant excess has a probability of 7.7$\cdot 10^{-8}$ (corresponding to
   5.4$\sigma$) of being caused by a background fluctuation. Correcting this
   probability with the appropriate trial factor, yields a 99.7\% confidence
   level (CL) for this excess to be related to the GRB (corresponding to
   2.7$\sigma$). This result is discussed within the framework of the WATCH
   detection.
   \end{abstract}
%
%
\section{Introduction}
Emission of TeV/PeV gamma-rays associated with GRBs has been extensively
searched. These studies are motivated by both the experimental results obtained
with the satellite experiments as e.g. EGRET and by theoretical models
(Meszaros et al. \cite{Meszaros}) which predict or at least allow TeV emission
to be produced in GRBs. So far none of these searches have revealed any
convincing evidence for VHE emission (see for example Aglietta et al.
\cite{Aglietta}; Alexandreas et al. \cite{Alexandreas94}; Borione et al.
\cite{Borione}; Connaughton et al. \cite{Connaughton}; Dazeley et al.
\cite{Dazeley}), although some tentative positive evidence has been found
recently in this energy range (Plunkett et al. \cite{Plunkett}; Krawczynski et
al. \cite{Krawczynski95}; Amenomori et al. \cite{Amenomori}).\\
Up to now five BATSE GRBs have been detected by EGRET with photons of energies
up to 18 GeV (Hurley et al. \cite{Hurley94}). Those GRBs are among the most
intense ones recorded by BATSE in the FOV of EGRET, so observations are
compatible with the hypothesis that all GRBs emit GeV photons but only the
strongest ones are above the EGRET sensitivity. A simple power law
extrapolation of the Superbowl GRB spectrum (Sommer et al. \cite{Sommer})
predicts that $\sim$20 photons above 20 TeV should be observed with the
extensive air shower (EAS) array of wide angle integrating Cherenkov counters
(AIROBICC) within 25 seconds while expecting only 0.1 background events. This
would lead to a highly significant detection by several other experiments
currently operating (Cherenkov telescopes, EAS arrays). The previous
extrapolation neglects source-intrinsic cutoffs and the attenuation through
interaction with the low energy cosmic photon background (Wdowczyk et al.
\cite{Wdowczyk}; Mannheim et al. \cite{Mannheim}; Stecker \& de Jager
\cite{Stecker}), which is expected for cosmological sources with redshift
greater than 0.1. Another interesting feature of strong GRBs is that their high
energy emission can be delayed and have longer durations than keV--MeV emission
(Hurley et al. \cite{Hurley94}). These ideas motivate searches for counterparts
at times and with durations independent of those given by the space detectors
at lower energies, especially when considering that the latter could miss GRBs
out of their FOV, or with a very hard spectrum, as has been already suggested
(Piran \& Narayan \cite{Piran}; see also Kommers et al. \cite{Kommers}).
Searches for GRBs with the HEGRA experiment using other data periods can be
found in Krawczynski (\cite{Krawczynski97}), Funk (\cite{Funk}), Padilla et al.
(\cite{Padilla97}), Krawczynski et al. (\cite{Krawczynski98}), Padilla
(\cite{Padilla98}).
\section{The AIROBICC array}
The AIROBICC array is part of the cosmic ray (CR) HEGRA detector complex
located at the Roque de los Muchachos on the Canary Island La Palma
(28.75$^\circ$ N, 17.89$^\circ$ W, 2240 m a.s.l.). It is an array of 7x7
stations (currently enlarged to 8x8 plus a subarray of 4x4 to increase detector
density in the center) with a regular grid spacing of 30 m, thus covering more
than 32000 m$^2$. Each station consists of a 40 cm diameter reflecting cone
which focuses the incoming light onto a fast 20 cm diameter photomultiplier
tube (PMT). The set is placed inside a protective hut with a lid which can be
opened through remote control. The PMT is covered with a blue filter ($\lambda=
$300-480 nm) to improve signal to noise ratio (S/N). The PMT output is
amplified in the hut and sent through a 150 m long cable to a constant fraction
discriminator (CFD) placed in the electronics container. The CFD is set to a
level equivalent to 5$\sigma$ of the night sky background fluctuations.
Whenever six or more AIROBICC stations exceed the CFD threshold within 200 ns,
a trigger is produced. Calibrations are performed every 20 minutes to measure
the relative delay between stations, the response of the TDC and the ADC
pedestals. The mean dead time after each recorded event is $\sim$8 ms. Detailed
descriptions of AIROBICC and the other components of the HEGRA complex can be
found in Karle et al. \cite{Karlea}, Fonseca et al. \cite{Fonsecaa}, Fonseca et
al. \cite{Fonsecab}.
   \begin{figure}
      \resizebox{\hsize}{!}{\includegraphics{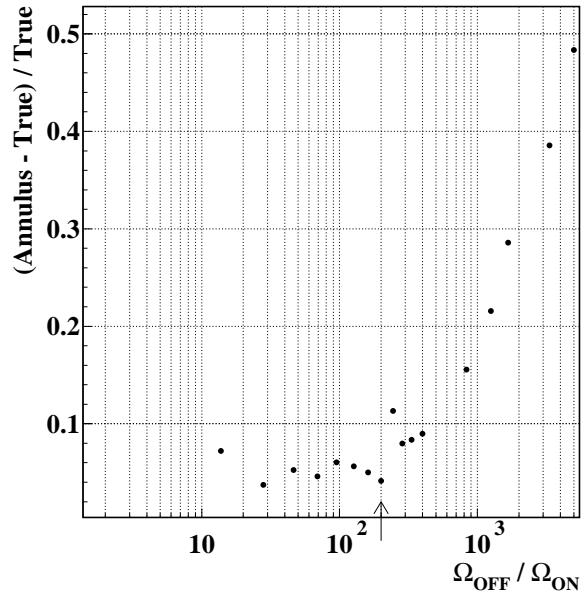}}
      \caption{Systematic error in the estimate of the background with the
         annulus method as a function of the ratio of the solid angles of the
         OFF and ON windows. The ON window is a fixed circle of 0.7$^\circ$
         radius and the OFF window is an annulus of fixed 1.5$^\circ$ inner
         radius and variable outer radius. The time interval is 4 minutes which
         is expected to have the lowest statistical error. The points are the
         average over several hundred windows. The sample used for this
         calculation consists of MC events generated as described in the text.
         The systematic error has a constant value of less than 5\% up to a
         certain size of the OFF window, from which it grows very fast. The
         arrow points to the value used in this work. This error has to be
         balanced against the statistical error which in the best case is 20\%.
         Note that the sign of the systematic error is positive, i.e. it is
         conservative.}
      \label{Error}
   \end{figure}\\
The ADC signals (AIROBICC + scintillator array) are used to locate the EAS core
position with a typical error $\le$18 m. The Cherenkov light front is fitted by
a cone to determine the arrival direction of the incident CR. The angular
resolution (angular distance containing 63\% of events for a point source) is
0.29$^\circ$ when 12 stations are fired (standard cut) and it continuously
improves with increasing number of triggered huts. The absolute pointing
accuracy has been estimated through comparison with the first HEGRA Cherenkov
telescope to be better than 0.2$^\circ$ (Karle et al. \cite{Karlea}). The FOV,
which is limited by the acceptance of the Winston cone, is 1 sr and the trigger
rate is $\sim$20 Hz. The energy threshold (50\% trigger probability) is
estimated to be $\sim$16(25) TeV for $\gamma$s and $\sim$29(37) TeV for hadrons
with zenith angle $\theta\le20^\circ(20^\circ<\theta<35^\circ$) using flux and
Monte Carlo studies (Mart\ic nez et al. \cite{Martinez}). The price that has to
be paid for the advantages of the Cherenkov technique (lower energy threshold,
better angular resolution) is that observations are restricted to moonless
clear nights ($\sim$10\% duty cycle).
\section{Analysis}
The data analyzed in this work were taken during the first year of operation of
AIROBICC, between March 1992 and March 1993. The mean trigger rate in this
period was $\sim$16 Hz. The sample contains $\sim4\cdot10^7$ events ($\sim$800
hours) and after a successful arrival direction reconstruction of those showers
with $\ge$7 fired stations, about $\sim2.5\cdot10^7$ events ($\sim$60\%)
survive. It has been shown that due to the low counting statistics encountered
when dealing with transient phenomena, one has to keep as many events as
possible; and so we do not apply the standard cut in the number of fired huts.
The angular resolution is then $\sim0.6^\circ$.\\
The analysis is based on a binned (in time and in space) all-sky search. We
define an ON window or source window and an OFF window or background window
where events are counted for a certain time interval. The search strategy
considers each event as being in the middle (in time) of a burst, so for every
single event in the data sample we take a pair of ON-OFF windows with a time
interval centered at the time of the event. The ON window is a circle of
0.7$^\circ$ radius centered at the position of the event and the OFF window is
an annulus of 1.5$^\circ$ and 10.0$^\circ$ radii concentric to the ON window.
The solid angles covered by the ON and OFF windows are denoted by $\Omega_{ON}$
and $\Omega_{OFF}$ respectively, and their ratio $\Omega_{ON}/\Omega_{OFF}$ by
$\alpha$. There is always one event inside the ON window and in order to not
overestimate the significance of a possible signal this event is not counted.
The size of the ON window is chosen to maximize the S/N according to our
resolution (see above) as described in Alexandreas et al. (\cite{Alexandreasa})
and it is expected to contain 75\% of the events for a point source. While the
size of the OFF window is taken as large as possible to increase statistics,
its limit is due to the systematic error introduced because of the non-linear
dependence of the counting rate on zenith angle (Alexandreas et al.
\cite{Alexandreasa}). We performed a Monte Carlo (MC) calculation to find the
largest radius that keeps the systematic error much lower than the statistical
error due to the low counting rate. Results are shown in Fig.~\ref{Error}.
The time intervals we chose are 10 seconds, 1 minute and 4 minutes which cover
the usual long duration GRBs. Dead time and low counting rate do not allow
searches for GRBs with much shorter duration.\\
The employed procedure has the advantage of being very sensitive but it is also
quite time consuming. So in order to look for TeV-emission on longer time
scales, a less sensitive search with a 1 hour time window and a classical
non-overlapping rectangular grid of 0.5$^\circ$x0.5$^\circ$ in celestial
coordinates has been carried out. In this case the background is estimated by
means of a MC method as described in Alexandreas et al. (\cite{Alexandreasa}).
For every real event we generate 100 MC events with random directions following
the acceptance function of AIROBICC (obtained with the events of a whole run,
therefore a different run implies a different acceptance function) but with the
same times as that of the original event. The number of MC events that fall
within the ON window determines the background.
   \begin{figure*}
      \resizebox{\hsize}{!}{\includegraphics{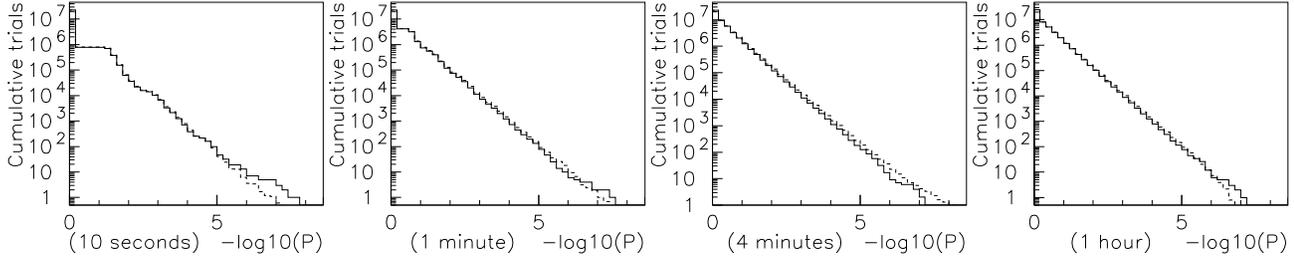}}
      \caption{Cumulative number of trials (windows) versus probability in the
         all-sky search. The time scale is indicated in brackets. The solid
         lines represent real data while the dashed lines represent the results
         (normalized to the real data) over a MC sample 10 (20 in the 1 hour
         scale) times larger than the real one. Deviations are not significant
         ($p\gtrsim 0.1$).}
      \label{Result}
   \end{figure*}\\
To obtain the significance for a hypothetical signal we use the probability
distribution given in Alexandreas et al. (\cite{Alexandreasa}) which is
appropriate for low statistics (Poissonian regime). It is the probability of
observing at least $N_{ON}$ events in the source window, given the observed
number of background events $N_{OFF}$, as a result of a background fluctuation:
   \[
      P(\geq N_{ON} \mid N_{OFF})
   \]
   \[
      = 1 - \sum_{n_{ON} = 0}^{N_{ON} - 1}\frac{\alpha^{n_{ON}}}{(1 +
      \alpha)^{n_{ON} + N_{OFF} + 1}}\frac{(n_{ON} +
      N_{OFF})!}{n_{ON}!~N_{OFF}!}
   \]
   \begin{table}
      \caption{90\% confidence level upper limits for the integral flux (in
         units of 10$^{-8}$ cm$^{-2}$ s$^{-1}$) of any hypothetical GRB
         occurred in our data sample. They are tabulated depending on the time
         window and the declination band. Uncertainties are $\sim$40\%.}
      \label{UpLim}
      \begin{tabular}{ccccc}  
         \hline
         Time scale & 10 sec & 1 min & 4 min & 1 hour \\
         \hline 
         $F_{UL}(E>$16TeV) & 4.5 & 3.5 & 2.5 & 0.35 \\
         ($9^\circ < \delta < 49^\circ$) & & & & \\
         $F_{UL}(E>$25TeV) & 2.8 & 2.2 & 1.6 & 0.32 \\
         ($-6^\circ < \delta < 9^\circ$, & & & &\\
          $49^\circ < \delta < 64^\circ$) & & & &\\
         \hline
      \end{tabular}
   \end{table}
where $N_{ON}$ and $N_{OFF}$ are the number of events in the source and
background windows respectively. In case of random distribution of the events
(i.e., no GRBs), a cumulative histogram of the number of trials with a chance
probability lower than $P$ as a function of -log$_{10}P$ should follow a
straight line with slope -1, which cuts the Y-axis at the height of the total
number of windows (trials). A strong GRB or several weaker GRBs should
therefore appear as a deviation from the line. The significance of a deviation
can be calculated knowing that theoretically every bin content follows a
binomial distribution. Actually this approach is only an approximation because
the probability distribution is discrete and there is an oversampling (i.e.,
search windows overlap and thus trials are not independent) in the search with
short time windows. Therefore we estimate the significance repeating the search
over a MC sample which is 10 times larger than the real one in the case of
short time windows and 20 times larger in the case of 1 hour search window.
\section{Results}
\begin{figure*}
      \resizebox{.983\hsize}{!}{\includegraphics{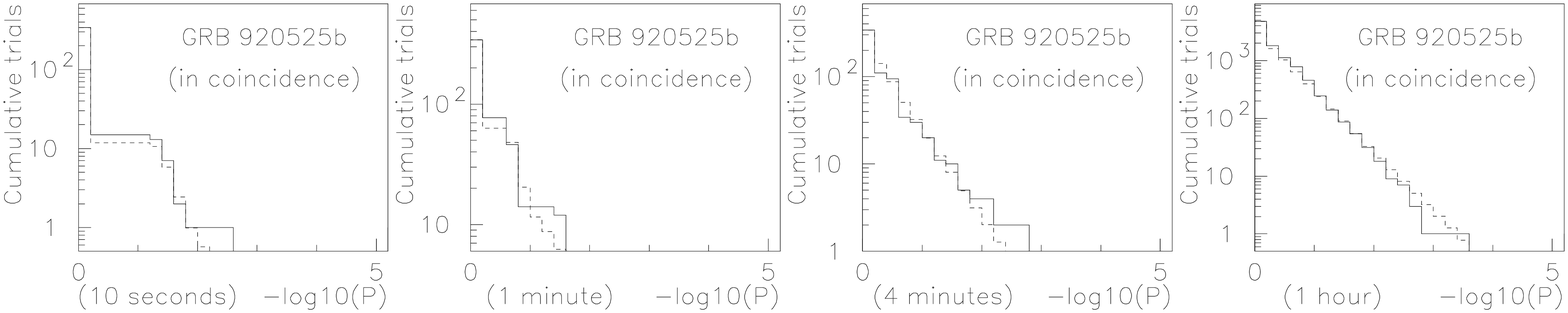}}
      \resizebox{.983\hsize}{!}{\includegraphics{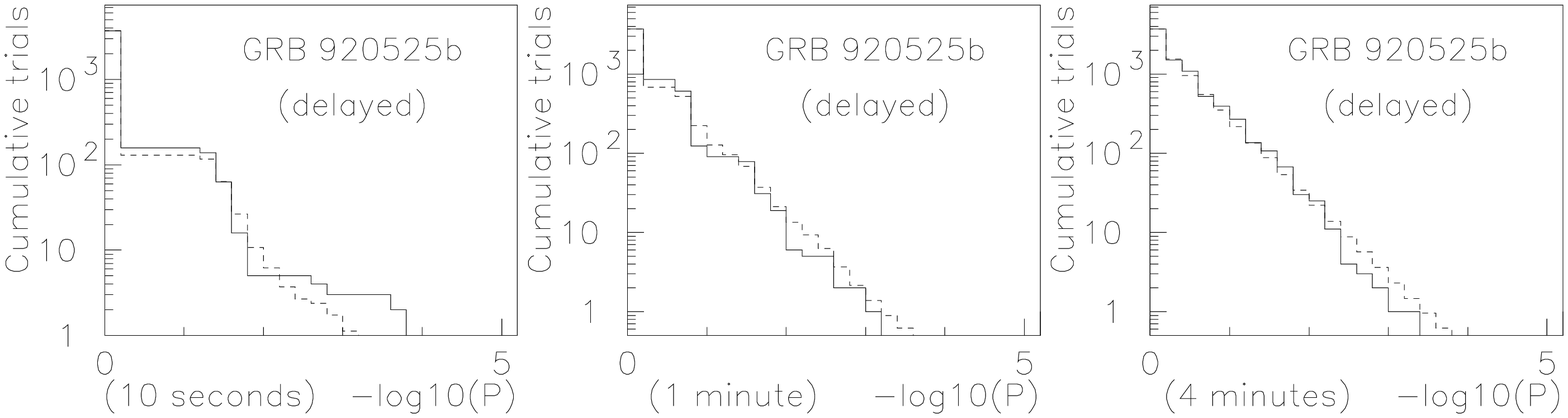}}
      \resizebox{.983\hsize}{!}{\includegraphics{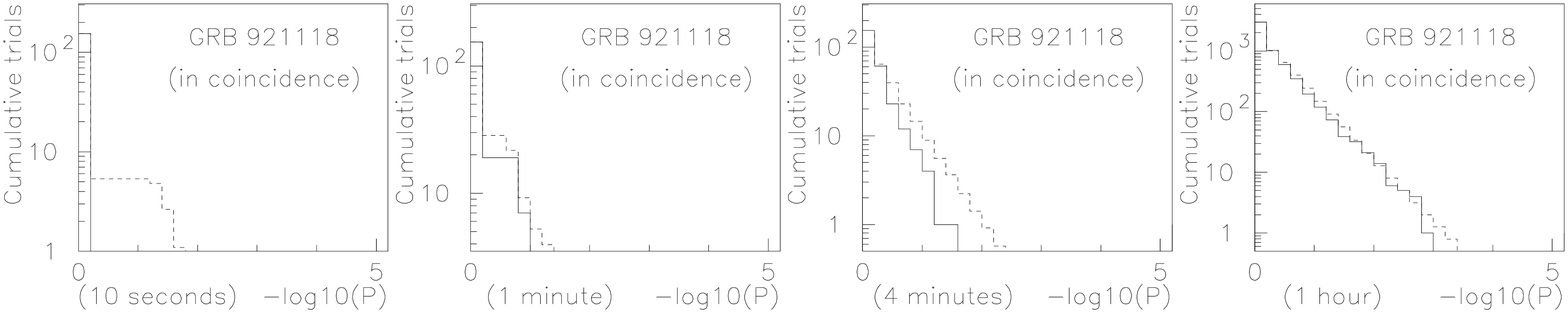}}
      \resizebox{.983\hsize}{!}{\includegraphics{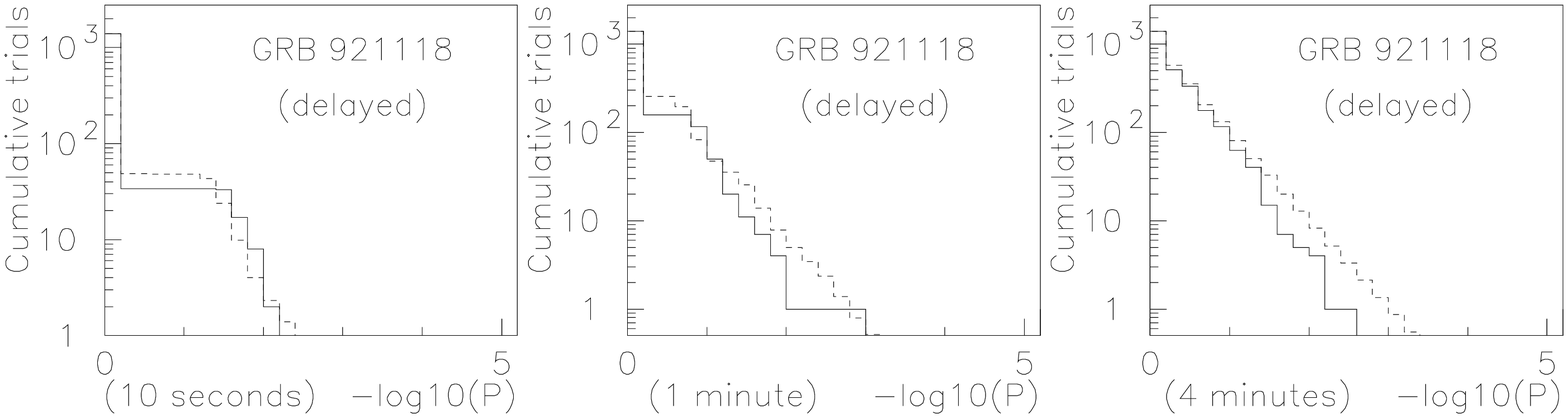}}
      \resizebox{.983\hsize}{!}{\includegraphics{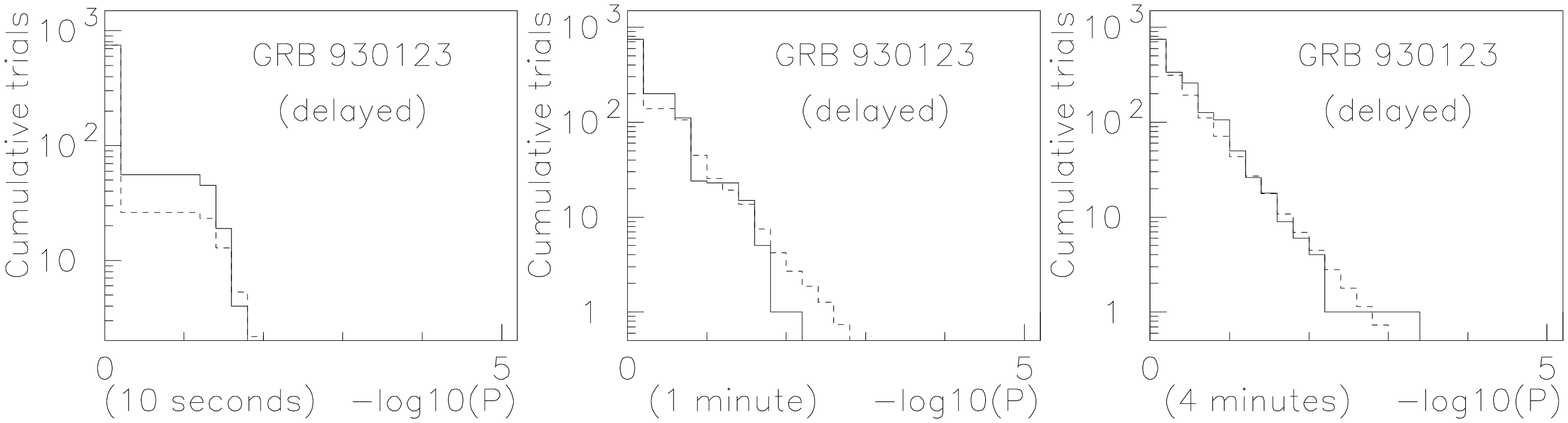}}
      \caption{Cumulative trials vs. probability in the search for the GRBs
         920525b, 921118 and 930123. We show the results of the search for
         coincident and delayed emission. For GRB 930123 only the result for
         delayed emission is shown because AIROBICC started more than one hour
         after the burst trigger. The dashed line shows the MC results. No
         significant deviation appears.}
      \label{GRBResult}
   \end{figure*}

The results for the all-sky search are shown in Fig.~\ref{Result}. It presents
the cumulative number of trials (search windows) against -log$_{10}P$ for the
all-sky search with the four time scales we have used. The results for the MC
sample are also shown. Obviously, the experimental data set is consistent with
MC expectations. The set of deviations has a chance probability greater than
$\sim$0.1. This result allows us to place an upper limit for the flux of any
hypothetical GRB which may have occurred in the FOV during our observations.
Depending on the declination band the resulting upper limit corresponds to two
energy thresholds for every time scale.\\
Assuming the same spectral index for the source and the CR flux and a steady
emission during the time interval, we can estimate the integral flux upper
limit with the formula (Karle et al. \cite{Karleb}; Alexandreas et al.
\cite{Alexandreasb})
   \[
      F_{UL}(E > E_{th_{\gamma}}) = \frac{N_{UL} \Omega_{ON}} {\alpha(N_{OFF} +
      1) \beta} F_{CR}(E > E_{th_{had}})
   \]
where $F_{UL}$ is the 90\% CL upper limit for the integral flux,
$E_{th_{\gamma}}$ and $E_{th_{had}}$ are the energy thresholds for $\gamma$ and
hadrons respectively, $N_{UL}$ is the 90\% CL upper limit for the number of
excess events in the source window as calculated by Aguilar-Ben\ic tez et al.
(\cite{Aguilar}) and $\beta$ is the fraction of source events expected to fall
in the source window. $F_{CR}$ is the known CR integral flux and is taken as
$1.8\cdot 10^{-5} E$(TeV)$^{-1.76 \pm 0.09}$ cm$^{-2}$ s$^{-1}$ sr$^{-1}$
(Alexandreas et al. \cite{Alexandreasb}; Burnett et al. \cite{Burnett}). We
applied this calculation to all ON-OFF windows and determined the highest
values. The resulting flux upper limits are listed in Table~\ref{UpLim}. Their
uncertainties are estimated to be $\sim$40\% through comparison with a
different method of flux calculation.
   \begin{table}
      \caption{List of GRBs observed in coincidence with satellites. $\theta$
         is the zenith angle of the GRB at the AIROBICC site.}
      \label{GRBs}
      \begin{tabular}{ccccc}
         \hline
         GRB & Observer & $\theta$ at trigger & $\theta$ 2$^h$ later & Coverage
         \\
         \hline 
         920525b & BATSE & 10.0$^\circ$ & 31.4$^\circ$ & Full \\
         920925c & WATCH & 5.1$^\circ$ & 25.1$^\circ$ & Full \\
         921118 & BATSE & 27.6$^\circ$ & 44.9$^\circ$ & Partial \\
         930123 & BATSE & 29.6$^\circ$ & 10.0$^\circ$ & Partial \\
         \hline
      \end{tabular}
   \end{table}\\
Additional constraints on the data sample as e.g. imposed by the burst
detections of BATSE, WATCH and other GRB detectors in space, can serve as a
tool to reduce the full data set and thus the expected statistical
fluctuations. Hence, the sensitivity is enhanced allowing detection of weaker
GRBs. We therefore searched in the BATSE 3B catalog (Meegan et al.
\cite{Meegan}) and in the WATCH catalog (Castro-Tirado \cite{Castro}; Sazonov
et al. \cite{Sazonov}) for triggers which were within the FOV of AIROBICC at
the time they occurred or up to 2 hours later (because of possible delayed
emission). They are shown in Table~\ref{GRBs}. The sample is then reduced to
events within $\pm10^\circ$ and $\pm$3 minutes around GRB locations. We also
looked for a delayed component for two hours after the initial triggers. The
size of the search region in celestial coordinates has been chosen, a priori,
large enough to have the uncertainties of GRB locations into account.
The length in time of the search interval (for coincident and delayed emission)
has been taken according to the observations in the GeV energy range (Hurley et
al. \cite{Hurley94}).\\
The most significant excess in the full data set was found almost in
coincidence with WATCH GRB 920925c, but from a direction 9$^\circ$ away from
the most probable WATCH position. This observation is discussed in the next
section. The results for the other three GRBs are shown in
Fig.~\ref{GRBResult}. It presents the cumulative number of trials against
-log$_{10}P$ for the four time scales used in the two search strategies
(coincident and delayed emission). The results for the MC sample are shown as a
dashed line. No significant deviation is seen. This, again, allows us to place
an upper limit for the integral flux of these GRBs. However, we only do so for
GRB 920525b because it is the only one completely covered by the data set among
the three GRBs. The upper limits for the integral flux are calculated at 90\%
CL in the same way as shown before. The results for the coincident and delayed
emission with the four time scales used here are shown in Table~\ref{GRBUpLim}.
Their uncertainties are estimated to be $\sim$40\%.
   \begin{table}
      \caption{90\% confidence level upper limits for the integral flux (in
         units of 10$^{-10}$ cm$^{-2}$ s$^{-1}$) of GRB 920525b in different
         time scales and for two scenarios: coincident and delayed emission.
         Uncertainties are $\sim$40\%.}
      \label{GRBUpLim}
      \begin{tabular}{ccccc}
         \hline
         Time scale & 10 sec & 1 min & 4 min & 1 hour \\
         \hline 
         $F_{UL}(E>$16TeV) & 89 & 8.9 & 3.2 & 0.40 \\
         (coincident, $\pm$3 min) & & & & \\
         $F_{UL}(E>$16TeV) & 130 & 14 & 3.2 & --- \\
         (delayed, $<$2$^h$ later) & & & & \\
         \hline
      \end{tabular}
   \end{table}
   \begin{figure*}
      \resizebox{\hsize}{!}{\includegraphics{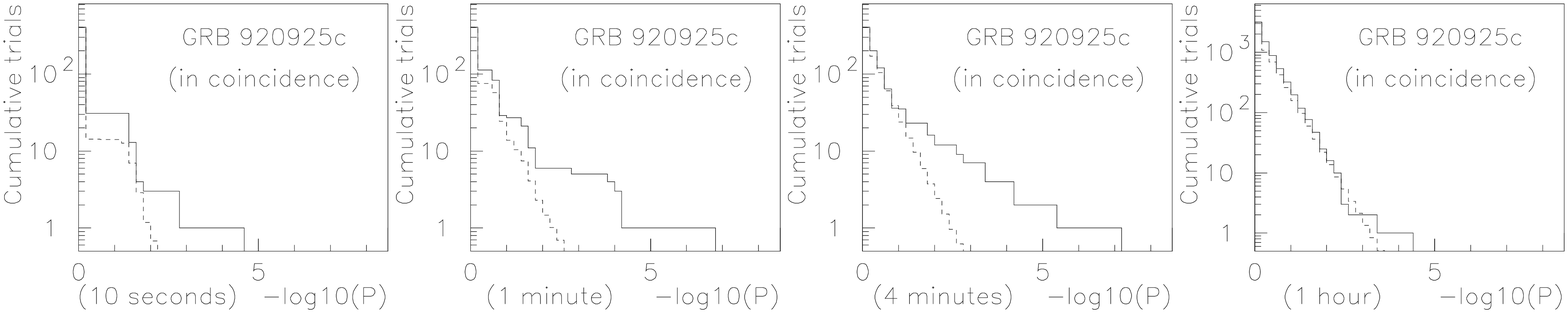}}
      \resizebox{\hsize}{!}{\includegraphics{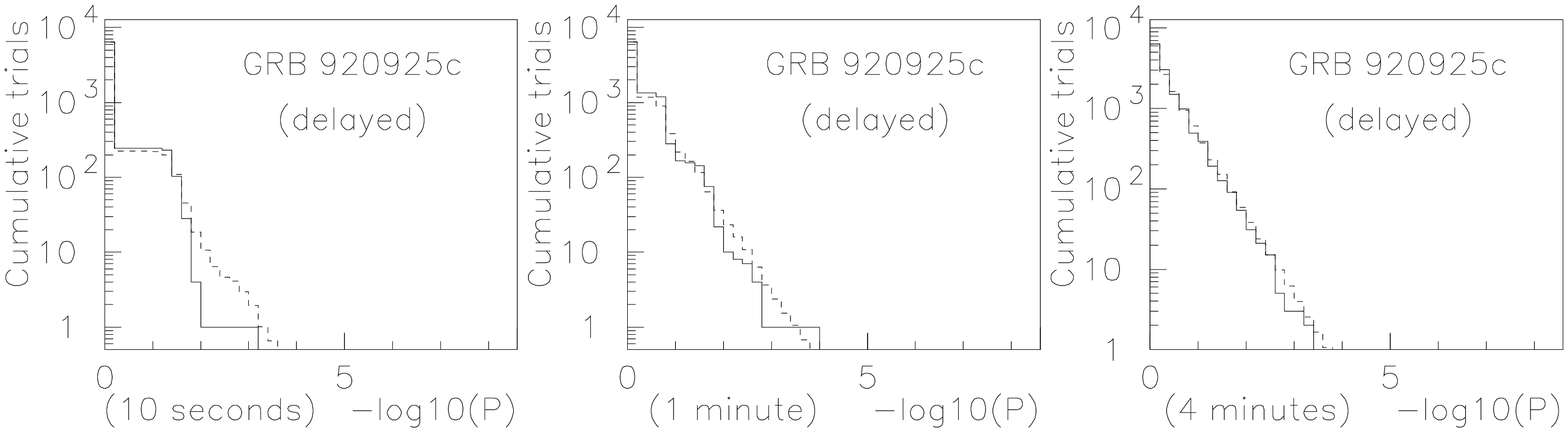}}
      \caption{Cumulative number of trials (windows) versus probability in the
         search for GRB 920925c. We show the results of the search for
         coincident and delayed emission. The dashed line shows the MC results.
         A significant deviation appears in some of the plots.}
      \label{WATCHResult}
   \end{figure*}
\section{Observational results for WATCH GRB 920925c}
\begin{figure}
      \resizebox{\hsize}{!}{\includegraphics{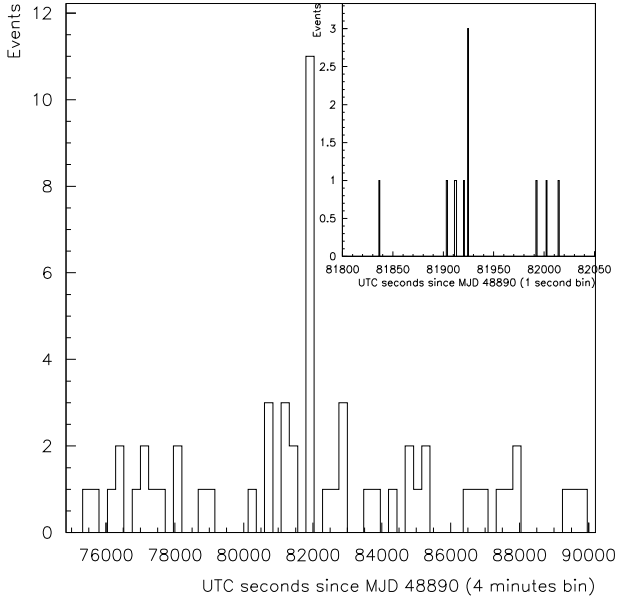}}
      \caption{Counting rate of events in the position of the excess detected
         by AIROBICC in coincidence with GRB 920925c along the four hours of
         moonless night. A clear peak is seen at the time of the excess, the
         inner figure shows in detail the time distribution of the burst
         events.}
      \label{Profile}
   \end{figure}

\begin{figure}
      \resizebox{\hsize}{!}{\includegraphics*[335,224][545,384]{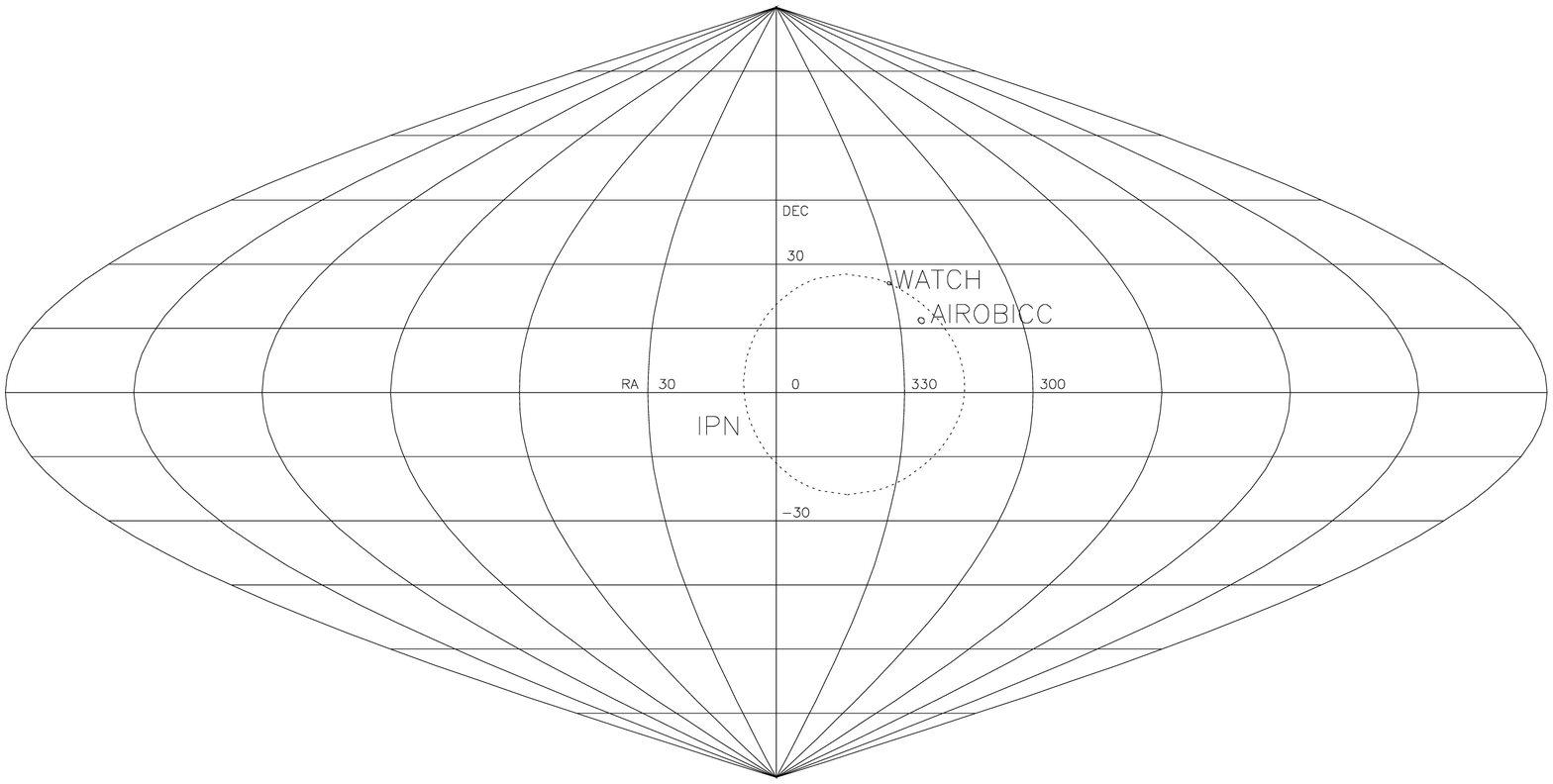}}
      \caption{Map with the situation of the WATCH GRB 920925c and the IPN
         ring as calculated from the WATCH and ULYSSES observations. The
         location of the excess detected by AIROBICC nearly coincident in time
         with WATCH is also shown.}
      \label{Map}
   \end{figure}

As it was mentioned in the previous section, the search for emission from GRB
920925c had a surprising result. An excess was found searching for emission in
coincidence with the WATCH trigger on all short time scales. No excess has been
found either in the search for delayed emission or in the 1 hour time window.
The results for the four time scales, in the two search modes (coincident and
delayed emission), are shown in Fig.~\ref{WATCHResult}. Again the cumulative
number of trials is plotted as a function of -log$_{10}P$ and the dashed line
shows the results for the MC sample. In this case significant deviations from
the expectations are observed, especially in the 4 minute time scale
(coincident emission), which shows a deviation with a chance probability
$<10^{-4}$.\\
The search interval with the smallest chance probability (4 minutes time scale,
coincident emission) is centered at $\alpha = 324.6^\circ \pm 0.3^\circ$,
$\delta = 16.8^\circ \pm 0.3^\circ$ (J2000) and UT = 22:45:21. Therefore it
precedes the WATCH trigger by less than one minute and is about 9$^\circ$ away
from the WATCH location. For this search interval we observe 11 events in 4
minutes while 0.93 events are expected. The chance probability for the
background to yield such an excess, computed according to the formula shown
above, is 7.7$\cdot 10^{-8}$ and the Li and Ma significance (Li \& Ma
\cite{Li}) is 5.4$\sigma$ (in both cases we take into account only 10 events
inside the search interval, see above). This is the most significant excess
seen in the whole data sample (compare Fig.~\ref{WATCHResult} with
Fig.~\ref{Result}). Correcting this probability with a trial factor (using the
MC) due to the search for the four GRBs in Table~\ref{GRBs} on several time
scales and in a large solid angle region, yields a final probability of
3.3$\cdot 10^{-3}$ (2.7$\sigma$), i.e. the CL for this excess to be related to
WATCH burst is 99.7$\%$ (neglecting the possibility of a second independent
burst). The data registered with the HEGRA scintillator array does not show any
excess, which may be due to its higher energy threshold and worse angular
resolution. The time distribution of the events registered by AIROBICC in the
position of the excess for the night of the GRB is plotted in
Fig.~\ref{Profile}. The distribution of the 11 events yielding the smallest
chance probability is shown in detail in the inner part of the figure. Note
that 7 out of the 11 burst events come within 22 seconds, 3 of which arrive
within 0.25 seconds.\\
Assuming that this excess is due to high energy emission we can calculate the
integral flux in the same way as the upper limits, but replacing $N_{UL}$ with
$N_{ON} - \alpha(N_{OFF} + 1)$. In this case the tentative mean integral flux
above 16 TeV during the 4 minutes window is (9 $\pm$ 4)$\cdot 10^{-10}$
cm$^{-2}$ s$^{-1}$.
   \\
The burst as seen by the space detectors had a duration of $\sim$5 minutes
exhibiting two main peaks. It was observed by WATCH and ULYSSES, thus reducing
the possible locations to an IPN (InterPlanetary Network) annulus (Hurley
\cite{Hurley96}). This annulus is obtained using the relative time of detection
of both spacecrafts. The IPN ring passes through the WATCH 3$\sigma$ error
circle and is 3$^\circ$ away from the excess observed with the AIROBICC array,
see Fig.~\ref{Map}.
   \\
Spectral data from WATCH in the interval 6-100 keV can be fitted by a power law
spectrum. The fit yields a spectral index of 2.5$\pm$0.2 which, if naively
extrapolated, does not predict any TeV flux detectable by AIROBICC. However,
the TeV spectrum may differ significantly from what one expects from the 100
keV extrapolation. Furthermore, the hypothetical emission at TeV energies
precedes the WATCH observations and may therefore be due to another production
mechanism.
\section{Discussion}
\begin{figure}
      \resizebox{\hsize}{!}{\includegraphics*[335,224][545,384]{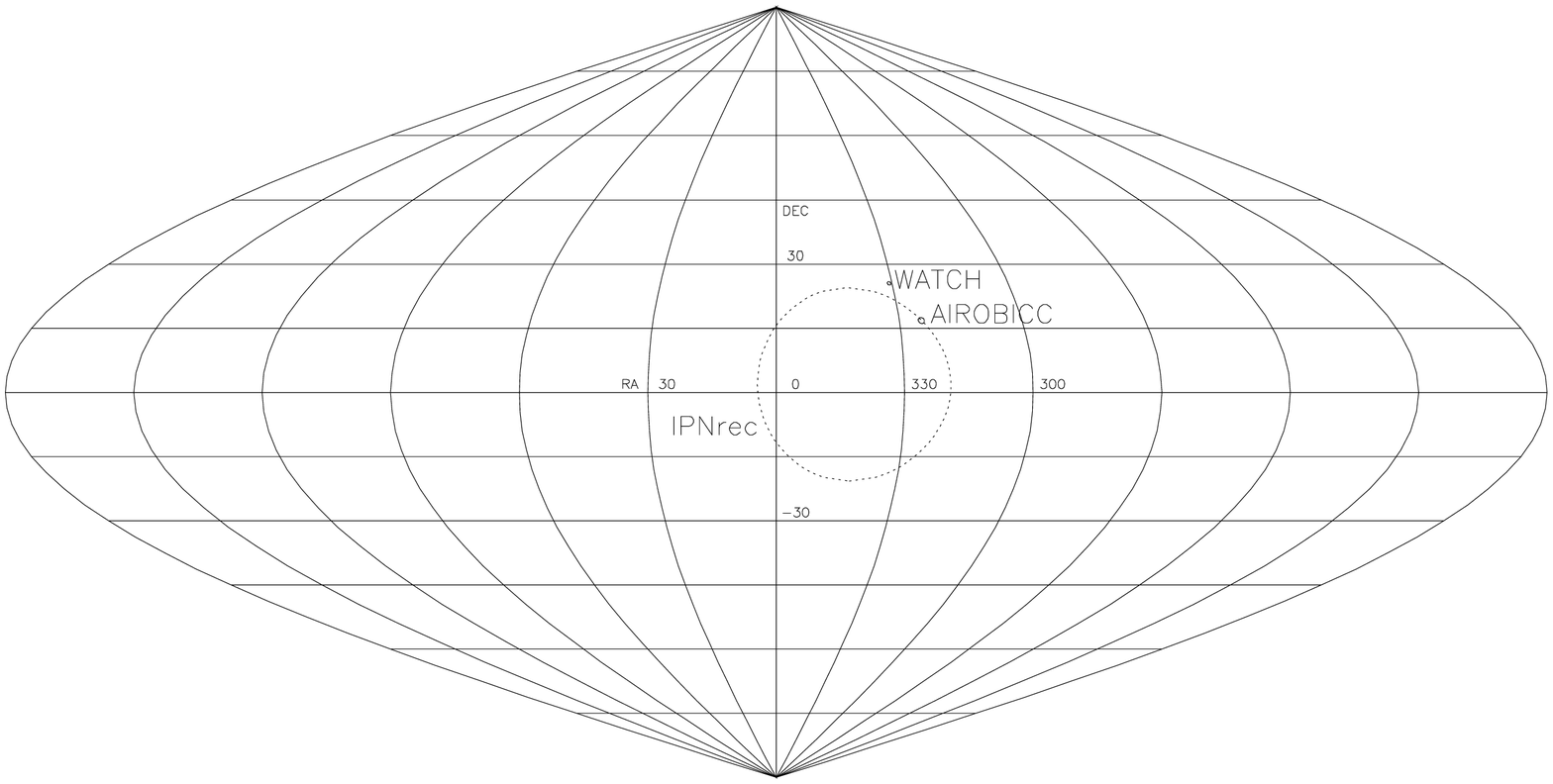}}
      \caption{Map with the situation of the WATCH GRB 920925c and the AIROBICC
       excess. The recalculated IPN ring using the ULYSSES and AIROBICC times
       of detection (see text) is also shown (IPNrec).}
      \label{Map2}
   \end{figure}

\begin{figure}
      \resizebox{\hsize}{!}{\includegraphics{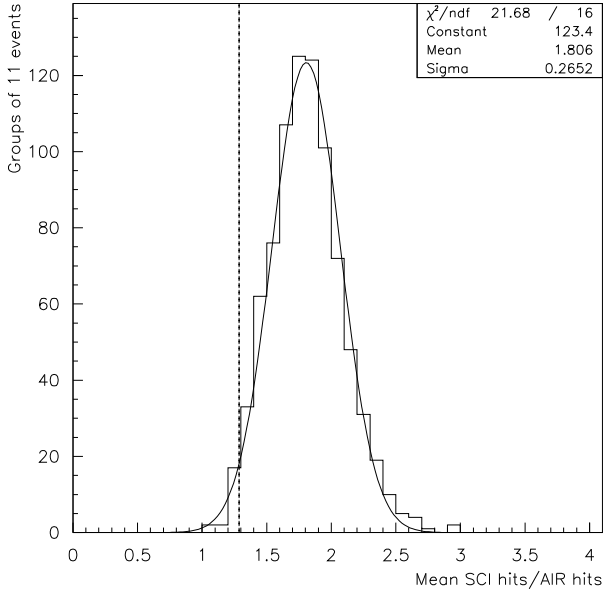}}
      \caption{Difference between the events of the excess and background in
       the ratio of fired huts in the scintillator and AIROBICC arrays for
       groups of 11 events. The mean for the 11 events of the excess (vertical
       dashed line) is below the mean for background groups (Gaussian fit) at a
       level of 2.0$\sigma$. The implications of this result are discussed in
       the text.}
      \label{Separa}
   \end{figure}

In order to add new clues that may clarify whether the excess recorded with
AIROBICC is related to GRB 920925c or not, the BATSE team (Kippen
\cite{Kippen}) investigated the reason why BATSE was not triggered by this GRB.
There are two reasons why BATSE may have missed the burst: either it was not
strong enough to overwrite the previous trigger (happened about one hour
before) or the Earth (or the Moon) occulted the source. The BATSE team kindly
provided us with a map of the sky, as seen by BATSE (Kippen \cite{Kippen}), at
the time that GRB 920925c occurred. It shows that the AIROBICC location was
completely occulted by the Earth throughout the whole burst. Therefore the
AIROBICC observation is compatible with BATSE not seeing the GRB. The WATCH
position was on the Earth's limb at the beginning of the burst and was occulted
about one minute later. The first peak of GRB 920925c (which is within the
first minute after the initial trigger) had lower peak flux than the previous
BATSE trigger and thus it would have not produced a trigger overwrite even
neglecting the additional attenuation by the Earth's atmosphere. So the
position given by WATCH is also consistent with the non-detection of the GRB
with BATSE.\\
A cross-check of burst positions given by BATSE, WATCH and IPN since 1991 to
1994 reveals disagreements for some events at a level of 5$\sigma$ or even
more, so an error in the WATCH location cannot be discarded (if the error is
due to timing it would also affect the IPN location estimate). Furthermore it
is interesting to note that assuming that the peak registered with AIROBICC
corresponds to the peak which triggered ULYSSES, we can recompute the IPN ring
using the time of detection of AIROBICC (neglecting the WATCH-AIROBICC distance
compared to the ULYSSES-Earth distance) yielding a new IPN ring which is
compatible with AIROBICC but not with WATCH (see Fig.~\ref{Map2}).
   \\
At this moment it is not possible to decide the nature of the excess recorded
with AIROBICC. Note that this GRB had a large fluence at keV energies, but was
not very intense. The GRB occurred under a small zenith angle (12$^\circ$) and
could thus be studied with a low energy threshold of the AIROBICC array. The
events of the excess differ significantly from background events as shown in
Fig.~\ref{Separa}. The figure plots the mean of the ratio of scintillator fired
huts to AIROBICC fired huts for groups of 11 events. The histogram fitted by a
Gaussian function represents the groups of background events. The vertical
dashed line, which is at 2.0$\sigma$ as given by the Gaussian fit, represents
the mean for the 11 events of the excess. It has been shown (Arqueros et al.
\cite{Arqueros}) that the ratio particles to light for EAS at observation level
is sensitive to the chemical composition and the difference observed in
Fig.~\ref{Separa} points to a gamma origin of the excess. This is confirmed by
a MC study in which simulated EAS are analyzed in the same way as real data
(Mart\ic nez et al. \cite{Martinez}; Cortina \cite{Cortina}). The results for
real data and for MC are summarized in Table~\ref{MCSepara}. If the excess has
indeed gamma-ray nature its significance would be higher due to the much lower
diffuse gamma-ray background compared to the charged particles background
(Karle et al. \cite{Karlec}). On the other hand the angular separation between
AIROBICC and WATCH locations is significant.
   \begin{table}
      \caption{Comparison between the ratio of scintillator fired huts to
       AIROBICC fired huts for groups of 11 events as predicted by the MC and
       as seen in the excess and in the background of the real data. The MC
       seems to confirm a gamma origin of the excess.}
      \label{MCSepara}
      \begin{tabular}{c|cc|cc}
         \hline
         $<$SCI/AIR$>$ & MC & & Data & \\
          & gammas & hadrons & excess & background \\
         \hline 
         Mean & 1.26 & 1.74 & 1.29 & 1.81 \\
         RMS & 0.17 & 0.23 & --- & 0.27 \\
         \hline
      \end{tabular}
   \end{table}
\begin{acknowledgements}
   The HEGRA Collaboration thanks the Instituto de Astrof\ic sica de Canarias
   for the use of the HEGRA site at the Roque de los Muchachos and its
   facilities. We also thank the BATSE, ULYSSES and WATCH teams for the
   availability of their data. We are especially grateful to A. J.
   Castro-Tirado, J. Gorosabel, K. Hurley, M. Kippen and S. Y. Sazonov. This
   work was supported by the BMBF, the DFG, the CICYT and a grant from the MEC.
\end{acknowledgements}
\end{document}